\documentclass[%
 aip,
 amsmath,amssymb,
 reprint,%
]{revtex4-1}

\usepackage{graphicx}
\usepackage{dcolumn}
\usepackage{bm}

\usepackage[utf8]{inputenc}
\usepackage[T1]{fontenc}
\usepackage{mathptmx}
\usepackage{etoolbox}

\makeatletter
\def\@email#1#2{%
 \endgroup
 \patchcmd{\titleblock@produce}
  {\frontmatter@RRAPformat}
  {\frontmatter@RRAPformat{\produce@RRAP{*#1\href{mailto:#2}{#2}}}\frontmatter@RRAPformat}
  {}{}
}%
\makeatother
\begin{document}

\preprint{AIP/123-QED}

\title[Contour Extraction of ICF Images]{Contour Extraction of Inertial Confinement Fusion Images By Data Augmentation}
\author{M. Falato}
\author{B. Wolfe}%
\author{N. T. T. Nguyen}
\author{X. Zhang}
\author{Z. Wang}
\thanks{The author to whom correspondence should be addressed: zwang@lanl.gov}
\affiliation{ 
Los Alamos National Laboratory, Los Alamos, NM, 87545, USA 
}%


\date{\today}

\begin{abstract}
X-Ray radiographs are one of the primary results from inertial confinement fusion (ICF) experiments. Issues such as scarcity of experimental data, high levels of noise in the data, lack of ground truth data, and low resolution of data limit the use of machine/deep learning for automated analysis of radiographs.  
In this work we combat these roadblocks to the use of machine learning by creating a synthetic radiograph dataset resembling experimental radiographs. 
Accompanying each synthetic radiograph are corresponding contours of each capsule shell shape, which enables neural networks to train on the synthetic data for contour extraction and be applied to the experimental images.
Thus, we train an instance of the convolutional neural network U-Net to segment the shape of the outer shell capsule using the synthetic dataset, and we apply this instance of U-Net to a set of radiographs taken at the National Ignition Facility.
We show that the network extracted the outer shell shape of a small number of capsules as an initial demonstration of deep learning for automatic contour extraction of ICF images.
Future work may include extracting outer shells from all of the dataset, applying different kinds of neural networks, and extraction of inner shell contours as well.

\end{abstract}

\maketitle

\section{\label{sec:level1} Introduction}

Inertial confinement fusion (ICF) plays an important role in the development of fusion technology.
Given the recent National Ignition Facility (NIF) achievement of a burning plasma condition\cite{ignition, NIF}, interpreting the data from ICF experiments is of paramount importance for the progression of ignition capabilities.
While there exists a plethora of diagnostics taken during an ICF shot, the X-Ray radiographs provide unique insight into the shape of both the inner and outer shell of ICF capsules during the shot. 
Extracting the contour of a shell over the course of a shot can quantify the implosion and kinetic energy of the imploding shell, determine asymmetries in the shell shape, and extract instability information for the shell \cite{merrit, montgomery, Kline}, which can be used to further optimize the neutron and fusion yield. 

The difficulty in automated contour extraction of inner and outer shells from ICF images extends across several domains of data analysis. 
The X-Ray radiographs of capsules with favorable materials for ICF have high, irregular noise and low resolution, which prevents the use of classical edge detection methods in computer vision to ICF images.
Thus more manual contour extraction methods, such as lineout methods \cite{lineout}, are often used. 
While lineout methods can extract contours of images with low noise or regularly characterized noise, ICF images taken using materials more favorable for emission feature extremely irregular noise that can be difficult for traditional denoising methods. 

Naturally, machine learning approaches for contour extracting are a potential method to turn to for this type of task. 
However, the previously mentioned issues of high, irregular noise and low resolution are accompanied by additional issues of low abundance of data and lack of ground truth data for supervised training of neural networks. 

In this work we address these problems that prevent the use of supervised deep learning \cite{reviewofdeeplearning} to extract ICF radiograph contours by developing synthetic radiographs with corresponding ground truth contours. We then use a generative adversarial network \cite{gans_Overview, GoodfellowGAN} (GAN) to transfer the noise characteristics of ICF images to these synthetic radiographs. This allows a neural network, such as U-Net \cite{UNET, UNet_review}, to train to extract a mask (the shape of the shell) given the synthetic image-mask pairs. 
If the noise transfer is done well enough to put the synthetic radiographs in the same domain as the actual ICF images, then applying the trained network to the experimental images should be able to generally extract full contours without the need for fine-tuning a conventional algorithm. 
Noise reduction is found to be helpful for traditional as well as deep learning methods such as U-Net.

\section{Our Dataset of ICF Images}
The ICF image data includes 115 radiographs taken from NIF. 
These images are from six shots, and are retrieved as part of indirect drive ICF implosions. 
Preprocessing is applied through a pseudo-flatfield correction \cite{Brad}, which aims to adjust irregular lighting imbalance due to backlighting effects and allow the images to be more uniform. 
The shots include a mix of double shell and single shell shots with different driving laser energies, pulse types, hohlraum sizes, hohlraum fills, outer shell thicknesses, and inner shell thicknesses.  

\section{Automated Conventional Contour Detection on ICF Images}

ICF image contours can be analyzed using lineout methods, where a center of a radiograph is chosen and lineouts are used to detect the position of capsule shells in shots \cite{lineout}.
With a set of contour points, the contour can be decomposed into a sum of Legendre polynomials. 
This process is difficult to automate on a general ICF image due to the unique noise in the data. 
A method that uses pixel values that are directly affected and perturbed by the extreme noise and low resolution that are characteristic of the data is of concern due to lack of generality and reliability of results. 

Based on the traditional lineout methodology, we develop a simple algorithm that detects potential contour points. A pictorial explanation of the algorithm is shown in the top half of Figure 1, with a panel of examples shown in the bottom half of Figure 1. 
The algorithm consists of essentially three steps: find initial candidate points with horizontal lineouts, using the candidate points find the geometric center and transform the image to polar coordinates from the center, and compute new horizontal lineouts (equivalent to radial lineouts in polar coordinates) for the final proposed contour points. The algorithm features several parameters that must be manually selected. These parameters are: the range of the original image to compute horizontal lineouts, the spacing between lineouts, the stride about each lineout to average the intensity over, the size of the convolutional filter to convolve down the lineout for smoothing, and two thresholds to limit the minima considered and distance between maxima and minima considered. 
\begin{figure}
\includegraphics[scale = 0.95]{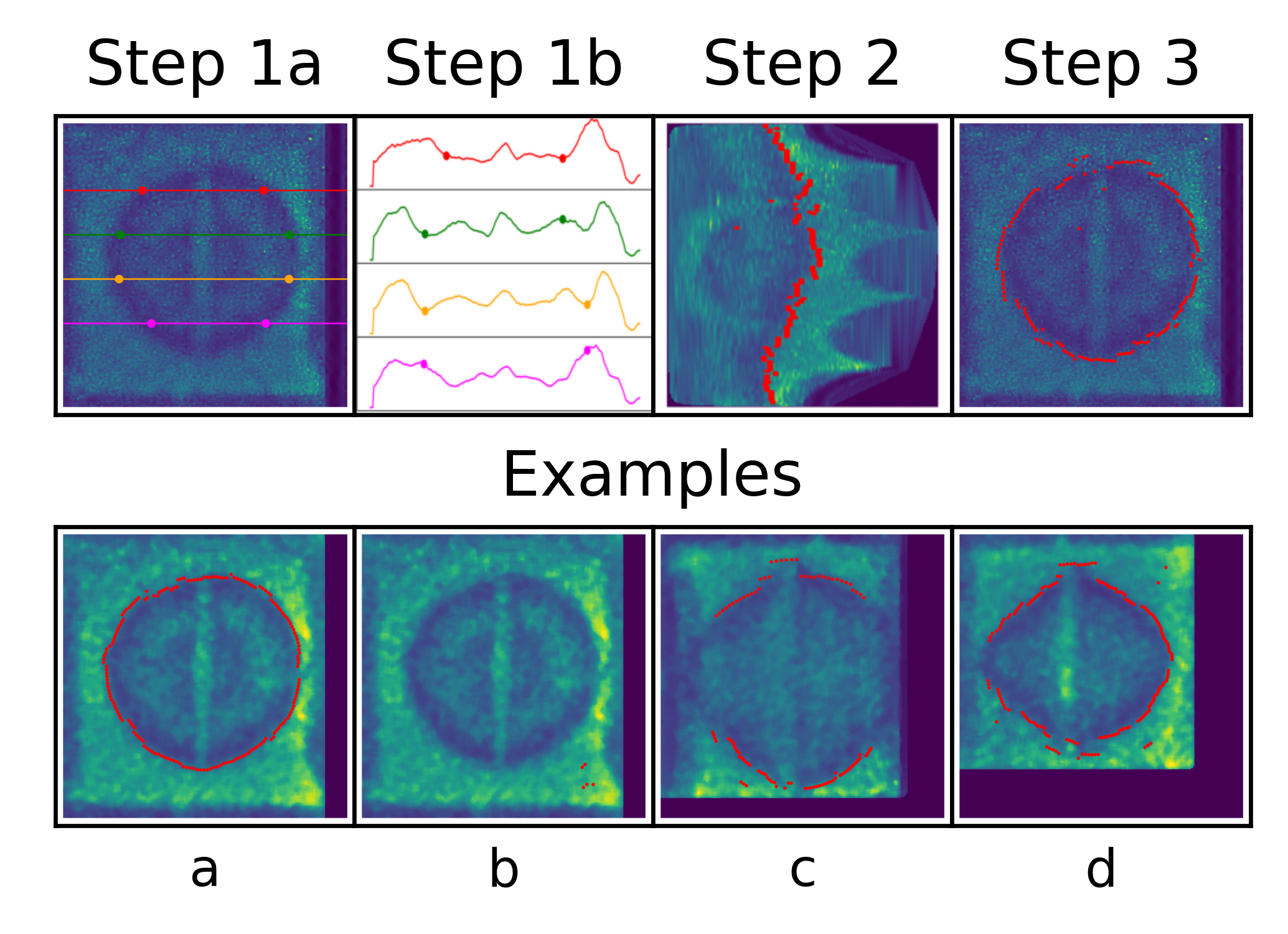}%
\caption{\label{} A representation of the algorithm in section III. Horizontal lineouts are used to detect initial contour points and a center of mass (Step 1). The image is transformed to polar coordinates about the center of mass, where horizontal lineouts from left to right are now equivalent to radial lineouts (Step 2), giving the final points. We then display the points in the original coordinates (Step 3). Only four lineouts are displayed in Step 1a to elucidate the lineouts and point selection in Step 1b. Examples are shown in the bottom panel, elucidating the problems with automating lineouts with a decent result (a), different parameterization failing to work on the same image (b), insufficient number of points (c), and need for manual removal of several points and postprocessing to recover a well-behaved contour (d). The best performance with the algorithm comes from images denoised from section V of this work.}%
\end{figure}

The amount of fine tuning required to discover a well-behaved contour is sensitive to the image noise and difficult to generalize. 
In general, one set of parameter values might find contours for part of the dataset while completely failing to in other parts of the dataset. 
We implement the algorithm on five different wavelet denoising instances of the full data set, pertaining to five different coefficient amounts and values for biorthogonal, decimated wavelet transforms. We also apply the algorithm to denoised images from section V of this work, which qualitatively perform the best with the algorithm. 
Over these runs, we do not find a parameterization suited for all images for Legendre decomposition. In order to extract any satisfactory Legendre decomposition, certain points must be hand eliminated or adjusted. 
The rest of this work describes contour analysis using deep learning, which has better automation and generalization potential. 
 
\section{Synthetic Data Generation}

In order to train a neural network for contour extraction, we first need a dataset with known contours, or ground truth. This is carried out in a two step process. First we create double shell capsules with physical characteristics, which are then made into synthetic radiographs via the Tomographic Iterative GPU-based Reconstruction Toolbox (TIGRE) \cite{TIGRE}. Image masks are created as the ground truth for the data. Then experimental-like noise characteristics are transferred to the synthetic radiographs using the Image-to-image Translation by Transformation Vector Learning GAN (TraVeLGAN) \cite{travelgan, brads_paper} in order to increase their resemblance to experimental data. 

Figure 2 displays a diagram of the workflow for the creation of synthetic radiographs. In order to produce data with contours commonly found in ICF we generate shells whose surfaces are defined using a sum of Legendre polynomials,
\begin{equation}
    r(\theta)=\sum_{n=0}^N a_nP_n(\cos (\theta))
\end{equation}
where in a 3D coordinate system $\theta$ is the polar angle. Azimuthal symmetry is assumed, where each contour can be rotated azimuthally to obtain surfaces. 
We generate shells with surfaces that have low mode asymmetries $(N<10)$.
During generation of the dataset we randomly sample the $a_n$ and reject selections where the surfaces self-intersect or intersect each other.
In the dataset we make the shells a uniform density defined by a specific material. 
For ablator shells we use aluminum and for the inner shells we use $SiO_2$ or germanium doped $SiO_2$.
With a 3D capsule, joint features and fill tubes can be added to the data by removing a portion of the shape.
The synthetic radiograph images are produced by projecting the objects using the Beer-Lambert Law,

\begin{equation}
    T := e^{-\int_L\mu(x)dl}
\end{equation}
where $T$ is the transmission, $\mu$ is the linear attenuation, $x$ is the 3D volume coordinate vector, and $L$ is the line between an X-Ray source point and a detector pixel.
We utilize TIGRE to numerically evaluate the above integral. 
TIGRE utilizes ray-tracing and allows easy adjustment of the source-object-detector geometry and can utilize GPU acceleration to speed up computation. 

\begin{figure}
\includegraphics[scale = 0.44]{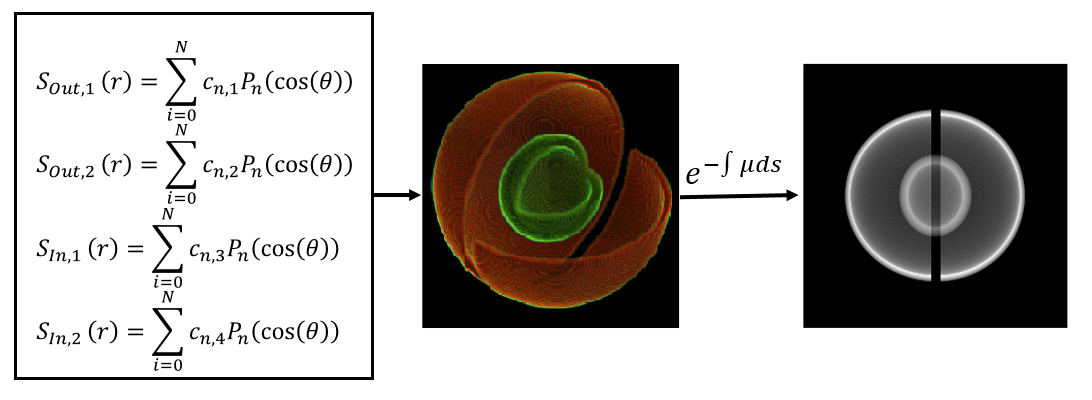}%
\caption{\label{} We generate four surfaces, $S_{out, 1}$, $S_{out,2}$, $S_{in,1}$, $S_{in,2}$ using a Legendre expansion of the radius values, $r$, and assume azimuthal symmetry. With four surfaces (two shells) we assign the inner shells and ablator shells the density of $SiO_2$ (or $SiO_2$ with a germanium dopant) and aluminum, respectively. We then emulate ideal radiographs by using the Tomographic Iterative GPU-based Reconstruction Toolbox (TIGRE).}%
\end{figure}

Since the experimental images have a low signal to noise ratio, it is important that the training data has similar noise characteristics.
We utilize the generative adversarial network, TraVeLGAN\cite{travelgan}, to apply noise to the noise-free synthetic images. 
Figure 3 displays a diagram of the workflow we use for the noise transfer on synthetic radiographs.
A generative adversarial network consists of a generator network and a discriminator network \cite{gans_Overview, GoodfellowGAN}.
The generator network is used to produce images that are similar to experimental radiographs.
The discriminator network is used to detect whether an image is an actual experimental image or an image produced by the generator.
By simultaneously training these two networks the system of networks produces images that resemble experimental images.

TraVeLGAN adds a third network to the typical system of generative adversarial networks, a siamese neural network \cite{SiameseNNoverview, siamesesignatureverification, siameseNNexplained}.
This network measures the similarity of images produced by the generator and the original projected images.
The generator is now not only required to generate images that trick the discriminator, but it must also generate images that have the same semantic features as the input data.
This makes it possible to generate images with experimental noise characteristics that have predefined contours.

\begin{figure}
\includegraphics[scale = 0.40]{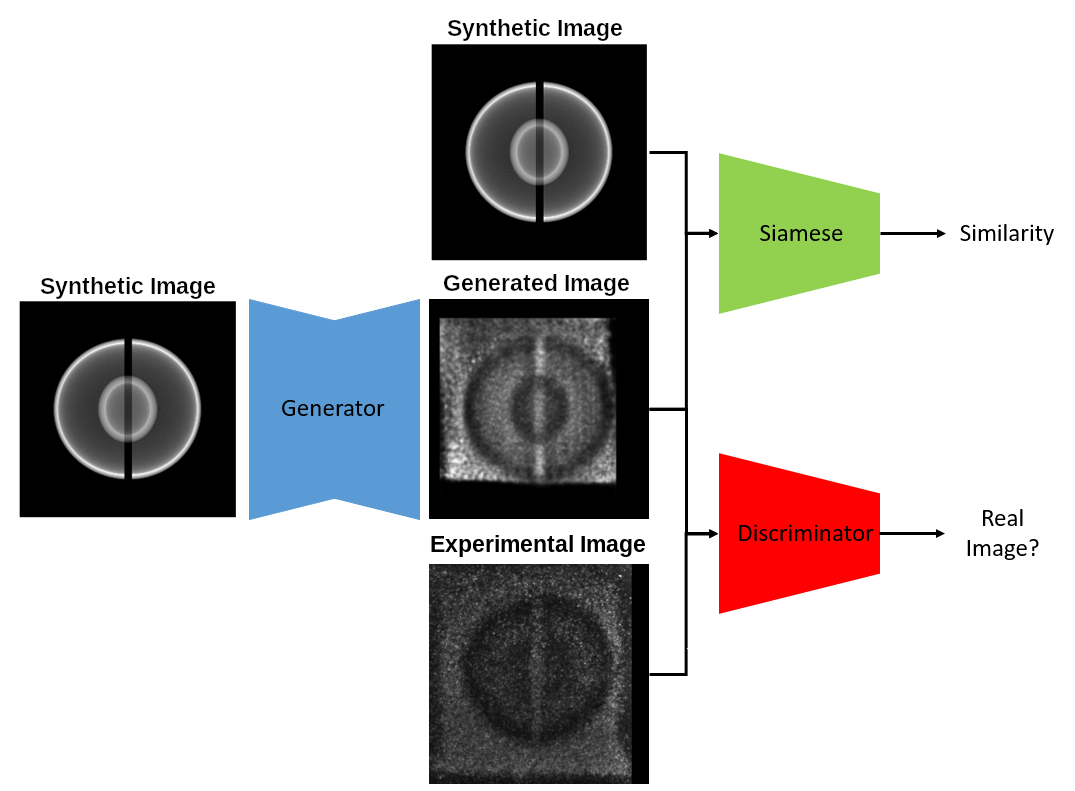}%
\caption{\label{} We use synthetic radiographs depicted in Figure 2 and transfer noise characteristics from experimental data to them. Image-to-image Translation by Transformation Vector Learning GAN (TraVeLGAN) is employed, where the generated noise characteristics are handled through the discriminator and the shapes in the generated images are handled through a siamese network coupled to the GAN.}%
\end{figure}

Using the techniques above, we furnish a dataset consisting of 2000 experimental-like radiographs. Accompanying each image, we create a mask image, an image showing the shape of the outer shell contour. We use the ground truth contours from the double shell creation to create an image of the outer shell shape, where we set pixels in the shape as 0 and pixels outside the shape as 1 for our masks. With image-mask pairs and a large data set, deep learning methods for contour/shape extraction can now be employed (see section VI).

\section{Denoising Techniques}
\begin{figure}
\includegraphics[scale = 0.3]{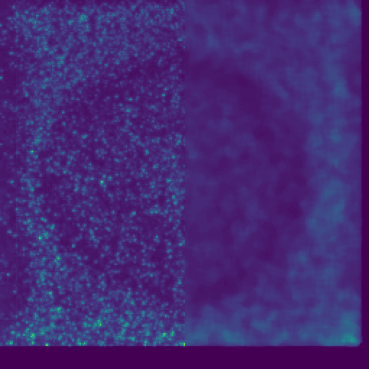}%
\caption{\label{fig:SnPdenoised} An example from the ICF image dataset (left half) containing salt-and-pepper-like noise after applying denoising protocol (right half) based on convolutional neural denoising network with or without ground truth.  }%
\end{figure}

Noise in corrupted inputs often adversely affects the image processing leading to much poorer accuracy performance of the study models\cite{Burger, Vincent}.  Image denoising by reconstructing clean inputs from noisy signals is proven to boost models' performances\cite{Khan,Tian,Nguyen}.  Convolutional neural networks \cite{reviewofdeeplearning} (CNNs) provide a strong platform to ``filter" image noise\cite{Khan,Dabov} due to i) the capability to learn features (including ``noisy" features) at different learning levels\cite{Goodfellow} via a stack of mathematical convolutional hidden layers and ii) the flexibility in designing suitable networks' depth and width at will.  We employ image denoising models built with convolutional layers inherited from DnCNN\cite{Vincent} and Noise2Noise\cite{Lehtinen} where the host neural network can be trained without clean inputs required. The underlying character of these types of denoising models has been shown\cite{Lehtinen}, where the converging fashion of a L2-norm cost function obtained with an U-Net-based\cite{UNET} (details in Sec. VI) denoising network is seen unchanged when the ground truths as targeted inputs, are replaced with another set of signals, i.e. in this case a different noisy ensemble.  Here, we customized the denoising parts\cite{Lehtinen} by adding a convolutional layer computing the salt-and-pepper noise content. We show an example result in Fig.~\ref{fig:SnPdenoised} 

The resulting denoised experimental images from this section are used in section III and VI of this work to increase performance of our traditional lineout algorithm and deep learning contour extraction, respectively.

\section{Deep Learning Method For Outer Shell Shape Extraction With U-Net}

Our synthetic dataset from section IV enables the use of deep learning on the original ICF dataset. We use the synthetic images to train an instance of the image segmentation CNN, U-Net \cite{UNET}. U-Net originated for the purpose of biomedical image segmentation, and it has been applied to CT scans, MRIs, X-Rays, microscopy, and many other forms of medical imaging\cite{UNet_review}. In general, U-Net trains on image-mask pairs in order to predict the correct mask, which is the exact type of data we are equipped with. The problem of contour extraction can be approached by performing edge detection on an output mask from U-Net. We implement U-Net using readily available code \footnote{https://github.com/milesial/Pytorch-UNet}.

\begin{figure}
\includegraphics[scale = 0.30]{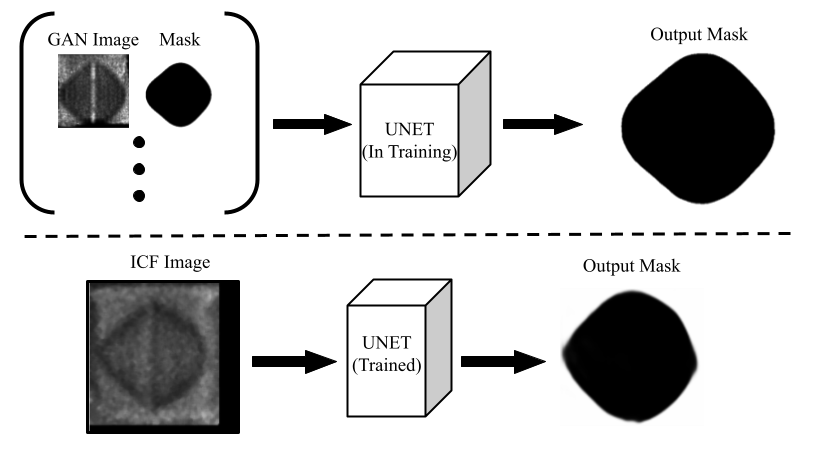}%
\caption{\label{} Diagram of the workflow using U-Net to segment the masks of outer shells in ICF images. First, we train the network on our synthetic data set (top half), then we apply the trained network to experimental ICF images (bottom half).}%
\end{figure}
U-Net consists of an encoder and decoder framework in a fully convolutional network. CNNs are commonly used for image classification tasks. While U-Net is a CNN, it does not classify an image as belonging to a class. Rather, U-Net classifies each individual pixel as being in the mask or not, and certainty for each pixel is converted to an output mask image.
Figure 5 displays the general workflow performed with U-Net once the synthetic dataset is provided. 
First we train an instance of U-Net on the image-mask pairs from our synthetic dataset (top half of Figure 5), which achieves a mean squared error loss of 0.0059 after 10 epochs showing the network sufficiently trains. 
After training is completed, we apply the trained network to experimental ICF images (bottom half of Figure 5), where 30 masks show complete contours. Because the outputs of the network are mask images, common edge detection and shape extraction methods can be used on output masks. The rest of the output masks either require postprocessing to fix shapes (see section VIII) or are beyond use.     

We use the marching squares algorithm\cite{marching_squares}, a common edge detection method, on the 30 masks showing complete contours. A panel of examples showing experimental ICF images, output masks from U-Net, and contours extracted is shown in Figure 6. 

\begin{figure}
\includegraphics[scale = 1.0]{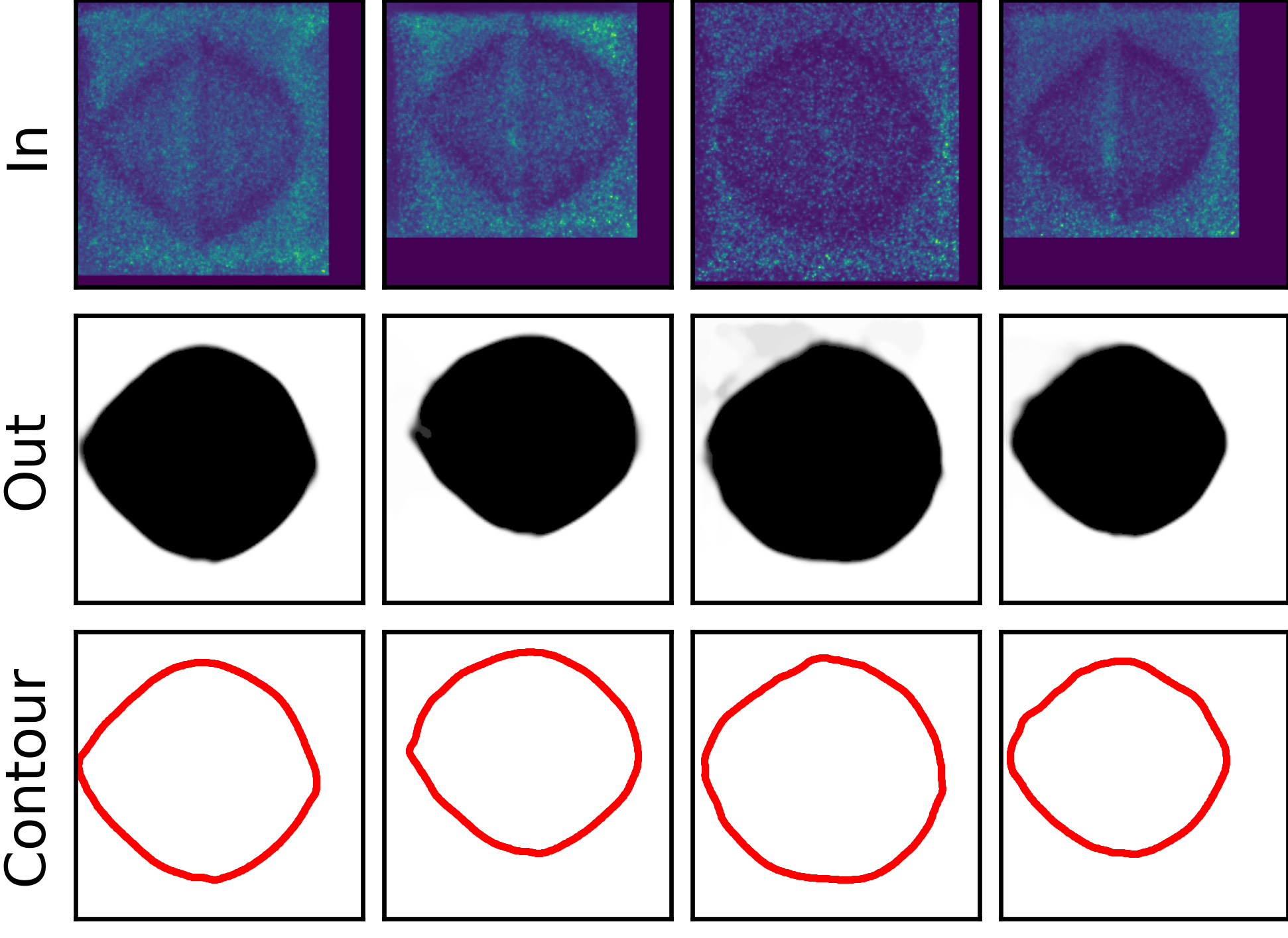}%
\caption{\label{} Panel of U-Net results showing the experimental ICF images (In), the output masks from U-Net (Out), and the contour extracted from the output mask using the marching squares algorithm (Contour). Note that ICF images are denoised using the method from section V before the network processes them.}%
\end{figure}

From the contours obtained, we conduct a Legendre fit and obtain a value for each coefficient (to fifth degree) in the expansion for each side of the shells. 
Distributions of each coefficient (divided by the base radius) for these results are approximated using kernel density estimation with Gaussian kernels with the sci-kit learn python package \cite{scikit-learn}. These distributions are displayed in Figure 7. We note that the even modes occupy broader distributions and odd modes are not as prominent. The range of even modes reasonably corresponds to experimental results  \cite{merrit}.  
\begin{figure}
\includegraphics[scale = 0.9]{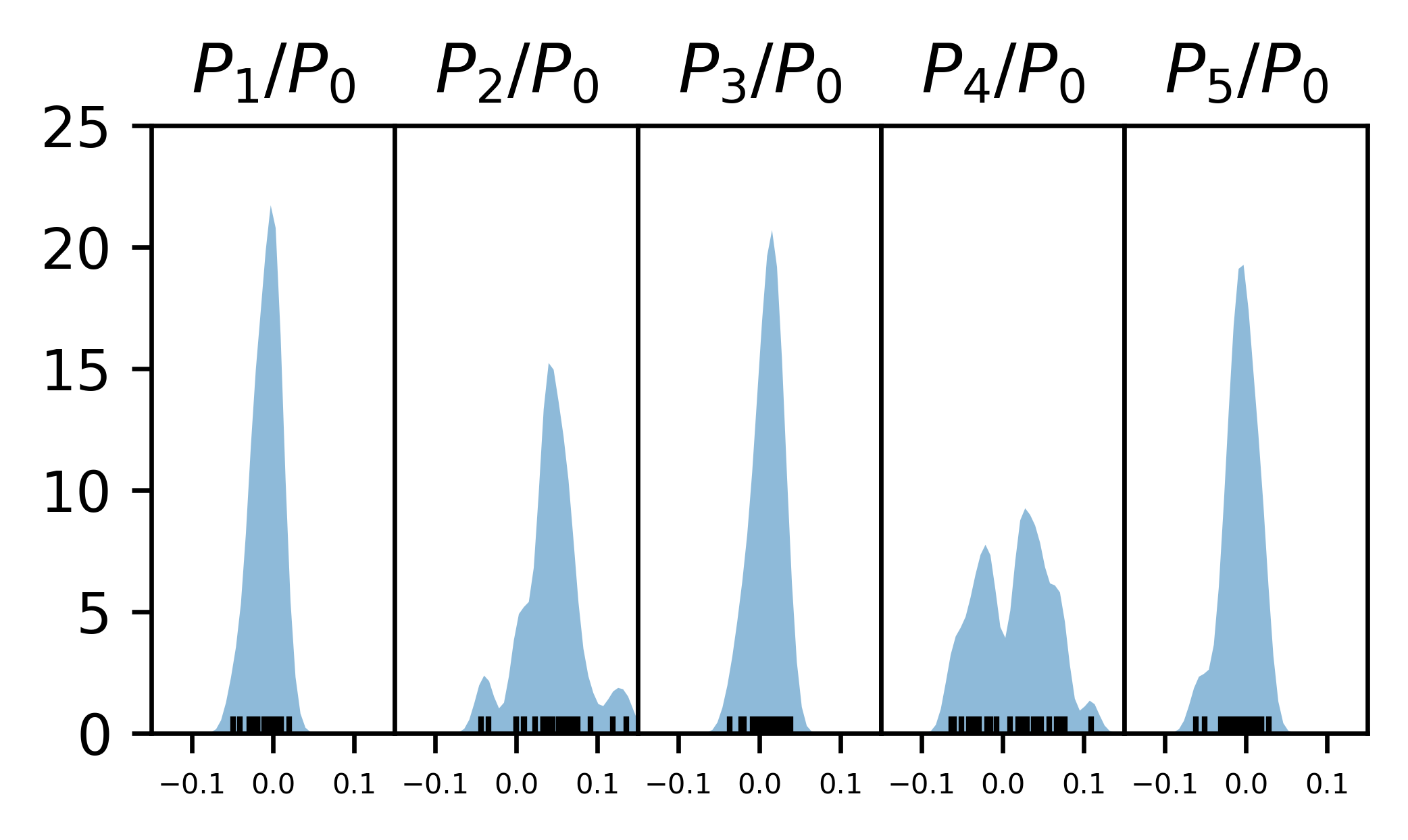}%
\caption{\label{} Estimated distributions of coefficients using kernel density estimation in the Legendre expansions from the contours extracted from U-Net. Ticks at the bottom of the distributions indicate the value of the mode a particular image has.}%
\end{figure}
A positive $P_4/P_0$ contribution corresponds to a more diamond-like shape, while a negative $P_4/P_0$ contribution corresponds to a square-like shape. To probe the results of the network, we can sort the experimental images by the values for each mode. We display images ranked by the value of $P_4$/$P_0$ mode coefficients in Figure 8, with the four highest valued and four lowest valued images obtained. These results agree with the notion of the $P_4$ polynomial in the Legendre expansion (1).
\begin{figure}
\includegraphics[scale = 1.35]{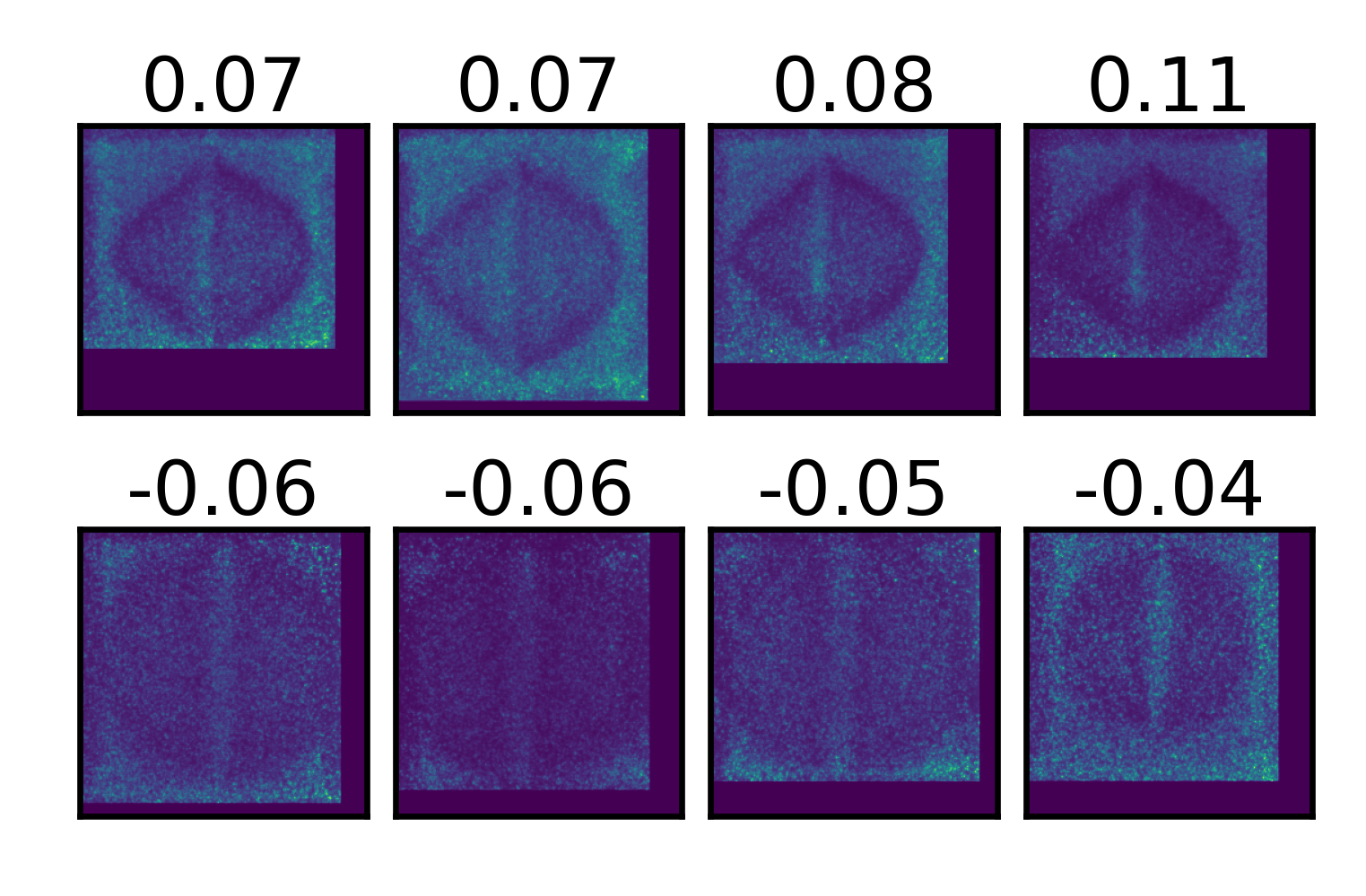}%
\caption{\label{} A panel of ICF images ranked by $P_4$/$P_0$ coefficient values obtained by the U-Net method. The top row displays the four highest values, while the bottom row displays the four lowest values.}%
\end{figure}

\section{Contour Completion}
\begin{figure}
\centering
\includegraphics[scale = 0.24]{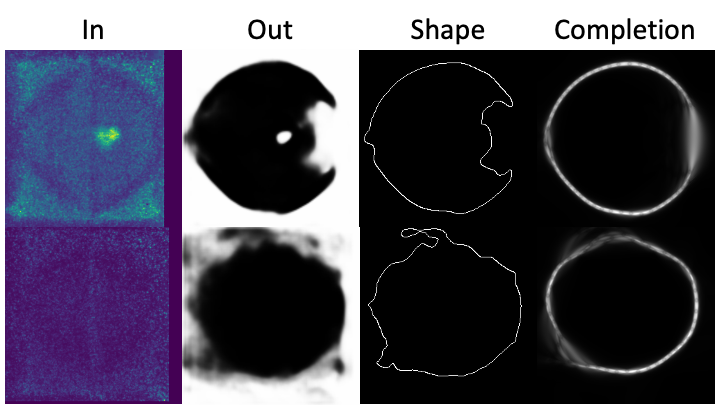}
\caption{The shapes of noisy masks. 
The contours in the right most column are computed from the contours in the second column from the right, with each row corresponding to one ICF image. 
The contour completion algorithm can fill in the missing contours, remove the irrelevant contours, and generate smooth contours.}
\label{fig:contourCompletion}
\end{figure}
Due to the low signal-to-noise ratio of ICF images, several resulting masks from U-Net are incomplete or have a noisy boundary. 
Accordingly, the extracted contours are noisy and the shapes of them are incorrect.
Figure \ref{fig:contourCompletion} illustrates this issue with two examples.
We address this issue by adopting a contour completion approach which is originally designed for perceptual completion of occluded surfaces \cite{Williams1996}. This approach has been used for detecting salient closed contours \cite{Williams1998, Williams2003, Zweck2004}
The idea of this approach is to model the distribution of contours as trajectories of particles undergoing Brownian motion\cite{Mumford1994, Williams1997_1}.
This idea implies a smooth contour because the change of particle direction is Gaussian distributed with zero mean.
Given a set of points sampled from the noisy contours, we compute a transition matrix, whose entries represent the probabilities of particles going from one point to others.
A closed contour corresponds to one of the eigenvectors of this transition matrix \cite{Williams2001}.
In particular, the smooth closed contours shown in Fig. \ref{fig:contourCompletion} correspond to the largest positive real eigenvalues of the transition matrices computed from the noisy contours. 
The fact that it is an eigenvector-eigenvalue problem means that only a part of the sample points is preserved, and this is why the approach can remove non-smooth contours.
The gaps between the preserved sample points can be filled in with smooth contours \cite{Williams1997_1}.
The contour completion algorithm is implemented as a recurrent similarity group convolutional neural network \cite{Zhang22}.

\section{Discussion and Conclusion}

In this work we use deep learning to automatically extract outer shell contours from experimental ICF images. 
A traditional lineout algorithm is developed in comparison and shows the difficulties of automating such methods. The problems elucidated are requirement of fine tuning of parameters and lack of generalizability. 
The extreme and irregular noise featured in ICF images limits the use of traditional methods in the above ways.
The use of deep learning for contour extraction is an attractive solution to these problems, as there is more automation due to a neural network doing the heavy lifting and more generalizability due to a data driven approach being performed. However, there is more to investigate on exactly how well TraVeLGAN learns the noise of the experimental data, which could explain the lack of sufficient results. 
Our contour extraction methodology is enabled through the use of a deep learning generative model on a set of synthetic radiographs. We perform this with only one network, however, while there exists several different forms of domain transfer networks\cite{st_review}. 
Furthermore, our methodology only uses one kind of image segmentation network out of the many that exist\cite{image_segmentation_survey}. Lastly, other types of networks, such as edge detection networks, could potentially be used for the problem. 

Future work includes implementation of more domain transfer neural networks for more data and different types of networks to extract the shape for different results and comparison of networks. Informing models using a modal error term in the loss might aid in the accuracy as well. We hypothesize that the improvement of synthetic data and network will increase the generalizability of the network and increase the amount of satisfactory results. Additionally, the ideas from this work can be applied to inner shell contour extraction as well. While there exists several improvements to be made, they are beyond the scope of this paper, where we introduce the architecture for this type of workflow for ICF images.  

The benefits of using our method include the automation of results (and thus, no need for fine-tuning of parameters), the use of the strengths of deep learning, and quick results once the network is trained. While the method only extracts contours on 25\% of the data, it is still a main priority to improve the generalizability of the method to more images in the data set so that it can be readily applied to new experimental data.

\section*{Funding}
This work was supported in part by the U.S. Department of Energy through the Los Alamos National Laboratory (ICF program). We thank the funding support from LDRD office, Mission Foundation project. Los Alamos National Laboratory is operated by Triad National Security, LLC, for the National Nuclear Security Administration of U.S. Department of Energy (Contract No. 89233218CNA000001). \\
\section*{References}


\bibliography{aipsamp}

\end{document}